# Stability of multivacancies in graphene

Ricardo Faccio[1,2] and Alvaro W. Mombrú[1,2,3]

1 Centro NanoMat, Cryssmat-Lab, DETEMA, Polo Tecnológico de Pando, Facultad de Química, Universidad de la República, Cno. Saravia s/n, CP 91000, Pando, Uruguay

2 Centro Interdisciplinario en Nanotecnología, Química y Física de Materiales, Espacio Interdisciplinario, Universidad de la República, Montevideo, Uruguay.

3 Author to whom all correspondence should be addressed

E-mail: amombru@fq.edu.uy

**Abstract**

The stability of graphene multivacancy systems is studied using Density Functional Theory (DFT) calculations. This work describes the evolution of the energy of formation per carbon atom for zigzag and armchair complementary figures -i.e. the figure formed by the carbon atoms extracted from graphene to form the vacancy-. Multivacancy systems formed when armchair complementary figures are removed are more stable for higher orders (>5) in comparison with the zigzag ones. The case of the construction of a 6-order vacancy from a 5-order one (branch-like) is discussed with the dependence on the place where the extra carbon atom is removed from graphene. The stability of multivacancy systems could be explained through the relative positions of the pentagonal rings present in the resulting defected graphene structure, as the more relevant factor. Other secondary factors that affect the stability of a graphene multivacancy system are the dangling bonds magnetic arrangements and their steric hindrance.

**Introduction**

Carbon materials have been intensively studied since the discovery of fullerenes [1] and the popularization of carbon nanotubes [2]. In the following years, graphite was studied [3-7] and more recently, research in graphene has become very interesting due to its very special physical properties [8]. The study of vacancies in nanotubes [9] and graphene [10,11] have been carried out due to the potential influence that such defects could cause on the pristine materials. Particularly, vacancies in graphene have been studied, either in mono- or divacancies [11,12] or in more complex multivacancies systems [13-17]. Recent reports describe the creation of defects by etching at high temperature and controlled atmosphere [18], and the improvement of the defects creation at the nanoscale in graphene [19], increasing the interest of theoretical work in this subject, to explain and to predict properties of new defected materials.

Recently, a set of rules that could help in the prediction of net magnetic moment in multivacancies in graphene, based on the shape of the complementary figure (i.e. the geometric figure of the atomic arrangement that is extracted from graphene when the multivacancy is created-) were reported [14].

The complementary figure gives a new insight to understand the multivacancies structures and their properties. For this reason we think that further work could be done in graphene multivacancies systems, regarding the different shapes that complementary figure can adopt.

In this manuscript we study the stability of simple multivacancy graphene systems by means of the first principles-based calculations of the formation energies of these defected systems. In order to carry out this study, we performed Density Functional Calculations, DFT [20,21].

Other systems with more complexity will be studied in other report [22].

It should be mentioned that the scope of this manuscript is disregarding the approach of reconstructed vacancies, 585 and 555-777 double vacancies or Stone Wales defects [23,24], just focusing in the vacancies formation without any rearrangement.

## 2. Computational Methods

Periodic spin polarized band structure calculations were performed on graphene multivacancies systems with the use of the DFT program VASP (Vienna *ab initio* simulation package) [25]; pseudopotentials were applied with a plane-wave basis. The exchange correlation potential was chosen as the generalized gradient approximation (GGA) [26] in a projector augmented wave (PAW) method [27,28]. An energy cutoff of 500 eV was used to expand the Kohn–Sham orbitals into plane wave basis sets. In all of the calculations reported here the k-point mesh was taken equivalent to 4x4x1 for the full (reducible) Brillouin Zone, allowing the convergence of total energy, stress components and ionic forces. Both supercell dimensions and ions positions were allowed to optimize, until residual forces and stress tensor components were positioned down to 0.01 eV/Å and 5 kbar respectively.

## 3. Discussion and Results

The formation energies per carbon atom, showing the stability in these multivacancy systems, were calculated in the following way:

$E(k) = E_k - NE_1$

where N is the number of carbon atoms in the graphene supercell after the vacancy has been created, k is the number of carbon atoms extracted to create the vacancy, $E_k$ is the energy of the optimized structure for a k-order vacancy and $E_1$ is the energy per carbon atom in regular graphene.

Figure 1 shows a scheme where the evolution of the complementary figure can be seen when the order of the vacancies (k) increases. The nomenclature $z$k and $a$k stands for zigzag and armchair growth of the complementary figure and $o$k for other configurations. Additional number n as in

*o*kn is just an ordinal number to identify different configurations among k-order vacancy systems.

As an example, figure 2 shows *z*4 (a) and *a*4 (b) optimized structures, with their respective complementary figures. The planar distortion of the lattice to form pentagons where terminal points of the complementary figures are is evident, as predicted previously [14].

The values of these formation energies per carbon atom for the *z*, *a* and *o* systems, are shown in Table 1.

Figure 3 shows a steady trend for increasing E(k) with the order of the vacancy, in armchair or zigzag complementary figures configurations. Linear fittings could be performed for these curves, both with acceptable $R^2$ = 0.96. Although there are some differences between calculations and the linear fit, it is possible to establish that for k > 5 the multivacancy systems built on the basis of an armchair complementary figure are more stable than the ones based on zigzag configurations.

However, a closer inspection for both curves shows that the increase between consecutive data is clearly dependent on the case. The energy differences E(k)-E(k-1) is smaller for even k than for odd k. In this way, both curves adopt a distorted staircase form. This effect could be seen in figure 4. Inset shows the E(k)-E(k-1) with the evolution of k. This figure is not very straightforward to explain. Black and hollow symbols stand for multivacancy systems related to zigzag and armchair complementary figure configurations, respectively. While the values of such differences is large for odd k -square symbols-, and low for even k values -diamond symbols-, the trend is to increase for armchair and to decrease for zigzag configuration, in all cases. These effects are shown by the lines and arrows in the inset. However, a very clear trend could be found for E(k)-E(k-2), where both curves correspond to either odd or even k values. This is shown in the main graph in figure 4: clear trends for the decrease and increase in the difference, for zigzag and armchair complementary figure configurations, respectively. It could be possible to deduce from both curves that at high order a convergence could be suggested. This convergence is represented by the dotted horizontal line, drawn just as a guide.

Figure 5 shows the cost to extract a carbon atom from the defected graphene system $o51$. In this figure, four different places where the vacancy could be enlarged are shown. The difference in energy indicates that $o64$ is the most stable vacancy, with $\Delta E = 0.416$ eV, followed by $o62$, $\Delta E = 1.852$ eV, and that $o63$ and $o61$ are the less stable ones, with $\Delta E = 2.099$ and $2.648$ eV, respectively. It should be remarked that this manuscript is restricted only to cases with dangling bonds. Other cases yield to specific research with multivacancy systems related to complex complementary figures and will be published soon [29].

This can be explained through the effects caused by the steric hindrance between the two dangling bonds that oppose the vertices of the complementary figure, according to the rules previously established [14], and the local ferro-, ferri- or antiferromagnetic configurations. This effect can be seen in figure 5. The possible ferro-, ferri- or antiferromagnetic configurations are easier to be determined through the shorter path of the complementary figure than through the longer path of the defected graphene itself. This is a secondary benefit of the use of the complementary figure when studying multivacancy systems. Thus, the $o63$ system is created from the $o51$ one, by extracting a carbon atom from the already defected graphene structure, with the occurrence of a second dangling bond, sterically perturbed by the original one, both in antiferromagnetic configuration. In comparison, the $o61$ system shows a less favoured ferromagnetic arrangement in a still sterically perturbed configuration. The $o62$ system shows no sterically perturbed dangling bonds configuration, but the ferromagnetic arrangement yields a non-favoured structure. The $o64$ structure is the more stable one in this figure, with no steric hindrance and antiferromagnetic arrangement.

Another example can be seen in figure 6, where the evolution of the vacancy -i.e. the complementary figure- from $a4$ to $o65$ through $o52$. While the $a4$ system has antiferromagnetic character, with steric hindrance, the $o52$ one is also sterically hindered but with ferrimagnetic arrangement for the three dangling bonds. According to the previous discussion, the $o$4 system should be more stable than the $o$52 one, which is not the case. The special feature for $o$52 and according to the rules already defined [14], the adjacent position of two pentagonal rings, which are located opposite to the terminal carbon atoms of the complementary figure, yields to a medium range low-stressed structure along the defected graphene system. This configuration, where the optimized structure is ruled by the competition among neighboring pentagonal rings, stabilizing the rest of the system seems to play a key role, more important than the other mentioned factors, i.e. the dangling bonds magnetic arrangement or their steric hindrance. When removing one more carbon atom and evolving to $o$65, an antiferromagnetic arrangement is formed but at the cost of the occurrence of three extra dangling bonds, and with no pentagonal rings to further stabilize the structure, yielding to a more unstable structure.

Figure 7 presents the partial density of states for $a$5, $z$5, $a$6 and $z$6. In all the cases the densities of states look asymmetric as consequence of the spin polarization, where σ- and π-states contributions are very different. The σ-states presents more localization due to presence of the multi atom vacancies, establishing a pseudo gap higher than 0.8 eV. In the case of π-states their contribution is present in a wider energy range, in comparison with σ-states, this is a consequence of the better $p_z$ hybridization between carbon atoms close to the edge of the multivacancy; which additionally exhibit an important contribution of states at the Fermi level. These results can be complemented by inspection of the corresponding spin-density maps, as presented in Figure 8.

**Conclusions**

Zigzag complementary figures, yield to extended vacancies, unstable when k > 5. Armchair complementary figures are more stable than zigzag ones, except for k = 4.

The stability of multivacancies could be explained through the relative positions of the pentagonal rings formed in the resulting defected graphene structure, as the more relevant factor. In fact the presence of two adjacent pentagons is suggested to yield an extra stabilization.

Other secondary factors that affect the stability of a graphene multivacancy system are the dangling bonds magnetic arrangements and their steric hindrance.

The complementary figure has proved again to be useful for the characterization of multivacancy systems and to help to predict their properties.

Further research on the stability of graphene multivacancy systems with other configurations is currently under analysis and will be published soon [29].

**Acknowledgments**

The authors wish to thank the Uruguayan funding institutions, CSIC, ANII and PEDECIBA.

| Vacancy system | E(k) (eV) | [E(k)-E(k-1)] / [E(k)-E(k-2)] (eV) | Vacancy system | E(k) (eV) | [E(k)-E(k-1)] / [E(k)-E(k-2)] (eV) |
|---|---|---|---|---|---|
| z3 | 8,194 | | | | |
| z4 | 10,793 | 2,599 | a4 | 14,103 | 5,909 |
| z5 | 16,154 | 5,361 / 7,960 | a5 | 15,330 | 1,227 |
| z6 | 17,800 | 1,646 / 7,007 | a6 | 15,744 | 0,414 / 1,641 |
| z7 | 21,878 | 4,078 / 5,573 | a7 | 18,473 | 2,729 / 3,143 |
| z8 | 22,727 | 0,849 / 4,927 | a8 | 19,314 | 0,841 / 3,570 |
| | | | a10 | 23,513 | - / 4.199 |

Table 1. Energy per carbon atom for defected graphene structures, differences with a single carbon atom extraction and with the extraction of two carbon atoms, for zigzag and armchair complementary figures.

**FIGURE CAPTIONS**

Figure 1. Complementary figures for graphene multivacancy systems. *z*, *a* and *o*, stands for zigzag, armchair and other configurations.

Figure 2.Structure of z4 and o4 configurations, with their complementary figures.

Figure 3. Evolution of the formation energy per carbon atom with the order of the multivacancy (black square symbols, armchair complementary figure; hollow circle symbols, zigzag complementary figure).

Figure 4. Differences in formation energy per carbon atom, among alternate order vacancy structures. Black square symbols stand for zigzag and hollow square ones for armchair (related to the complementary figure). The dashed line is guide to the eye for the trend for higher order vacancies. Inset: Differences in formation energy per carbon atom for succesive order vacancy structures. Black symbols for zigzag, hollow symbols for armchair. Square symbols for k odd, diamond symbols for k even. Arrows and dot lines are just a guide to the eye.

Figure 5. Effect of the position where a vacancy is grown from 5- to 6- order one. Large figures are E(*o*6n)-E(*o*51) in eV and the small figures below are the E(*o*6n) energies. The dangling bonds correspond to the vacancy opposed to the complementary figure -the one which is exhibited-. The defected graphene is omitted for clarity. Spin density configurations are shown by different colours.

Figure 6. Evolution of the vacancy from *a*4 to *o*65 through *o*52.

Figure 7. Partial density of states for: *z*5, *a*5, *z*6 and *a*6; indicating σ and π contributions. Blue and red lines represents π- and σ-states respectively.

Figure 8. Spin-density maps for *a*5, *z*5, *a*6 and *z*6, indicating majority and minority contribution in yellow and cyan respectively. All the graphs were obtained for an iso-density of 0.01 e/Å$^3$.

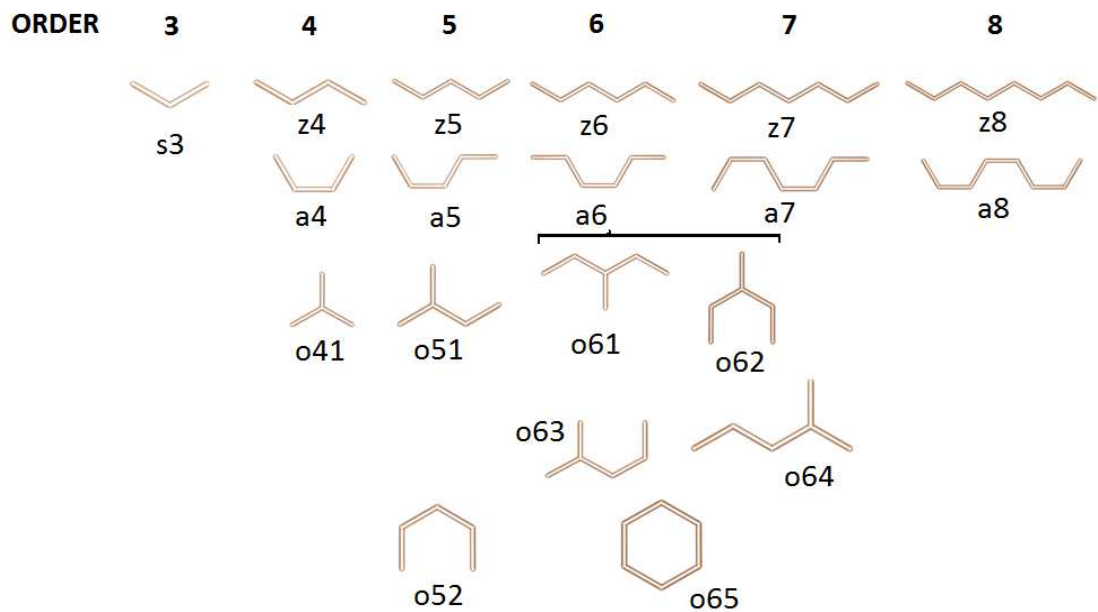

Figure 1. Complementary figures for graphene multivacancy systems. *z*, *a* and *o*, stands for zigzag, armchair and other configurations.

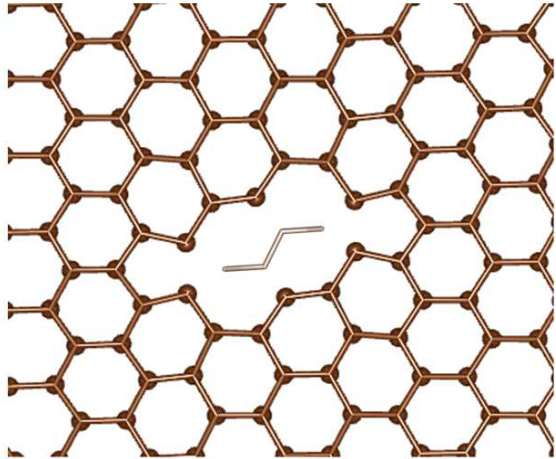

a)

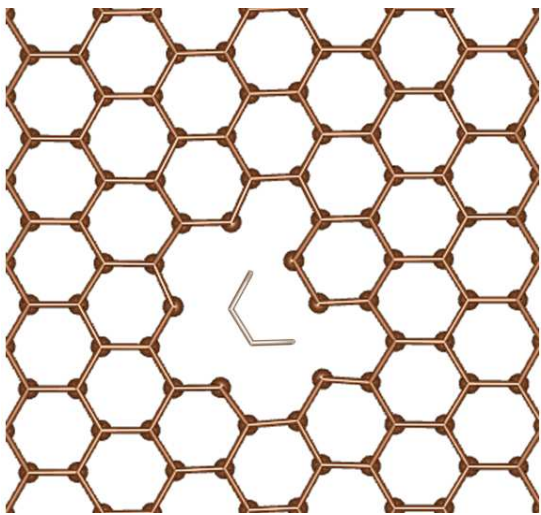

b)

Figure 2. Structure of z4 and o4 configurations, with their complementary figures.

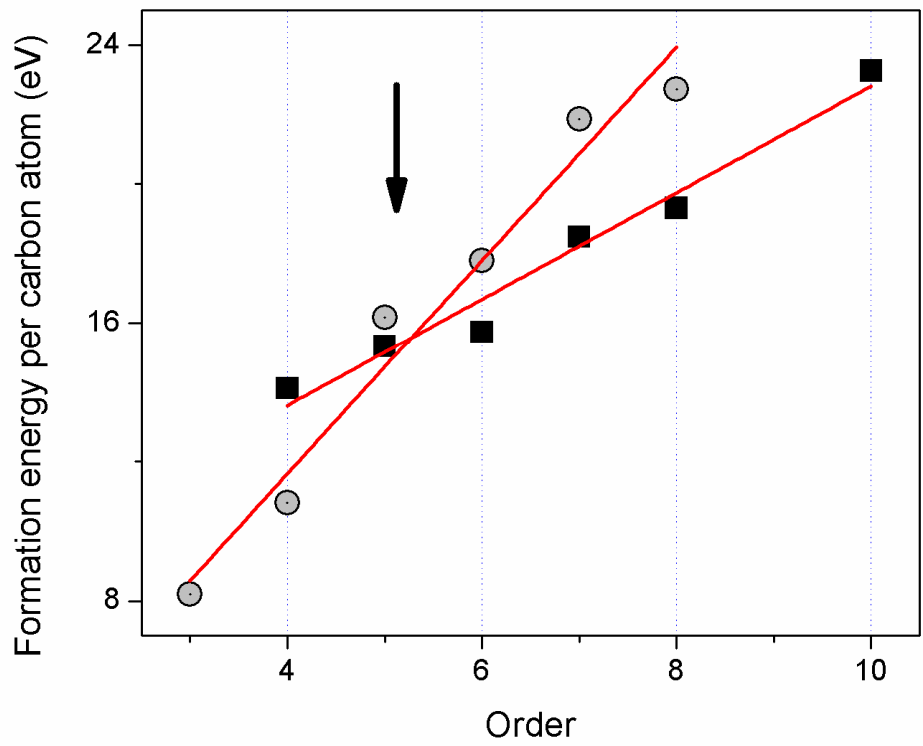

Figure 3. Evolution of the formation energy per carbon atom with the order of the multivacancy (black square symbols, armchair complementary figure; hollow circle symbols, zigzag complementary figure).

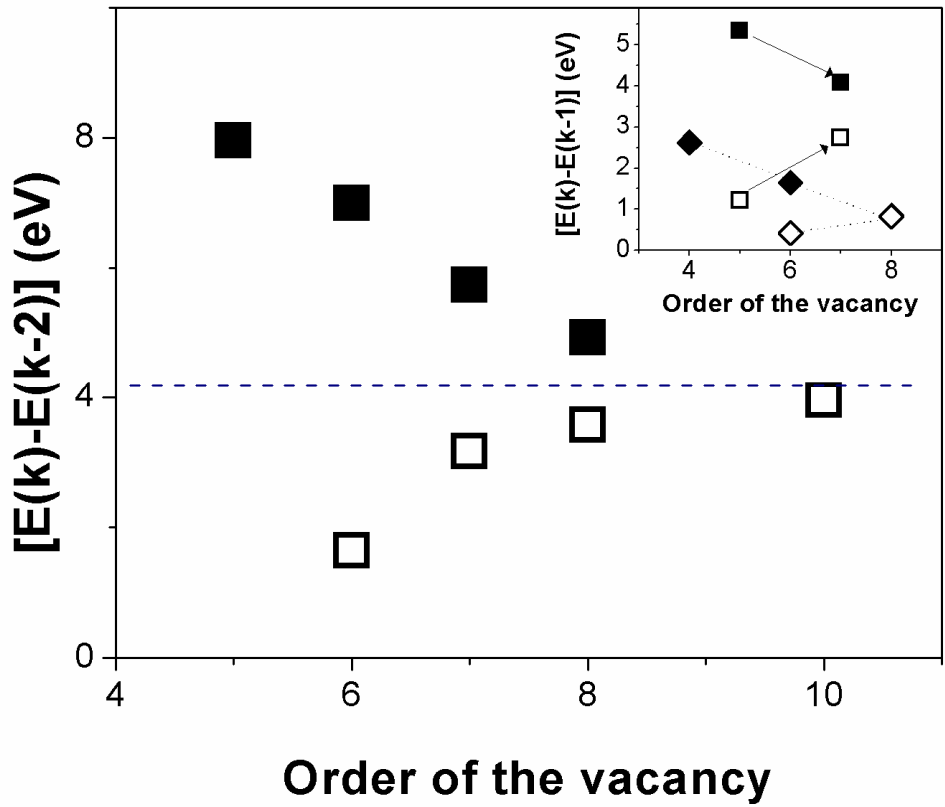

Figure 4. Differences in formation energy per carbon atom, among alternate order vacancy structures. Black square symbols stand for zigzag and hollow square ones for armchair (related to the complementary figure). The dashed line is guide to the eye for the trend for higher order vacancies. Inset: Differences in formation energy per carbon atom for succesive order vacancy structures. Black symbols for zigzag, hollow symbols for armchair. Square symbols for k odd, diamond symbols for k even. Arrows and dot lines are just a guide to the eye.

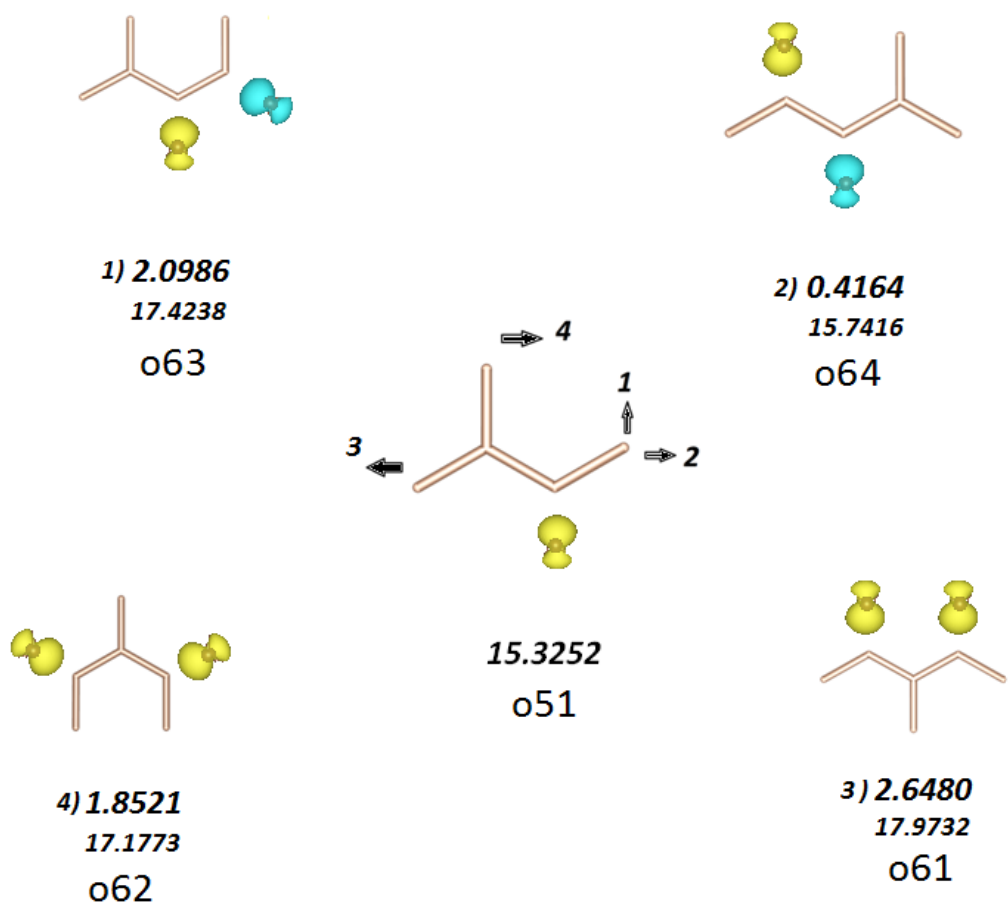

Figure 5. Effect of the position where a vacancy is grown from 5- to 6- order one. Large figures are E(*o*6n)-E(*o*51) in eV and the small figures below are the E(*o*6n) energies. The dangling bonds correspond to the vacancy opposed to the complementary figure -the one which is exhibited-. The defected graphene is omitted for clarity. Spin density configurations are shown by different colours.

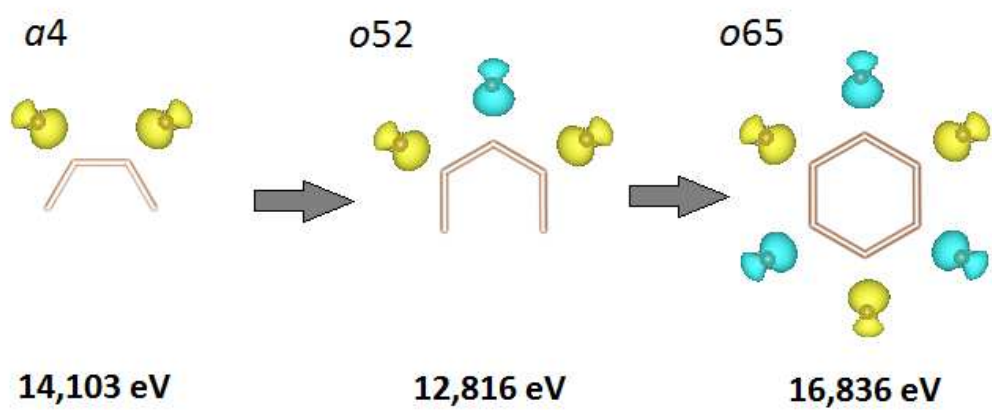

Figure 6. Evolution of the vacancy from *a*4 to *o*65 through *o*52.

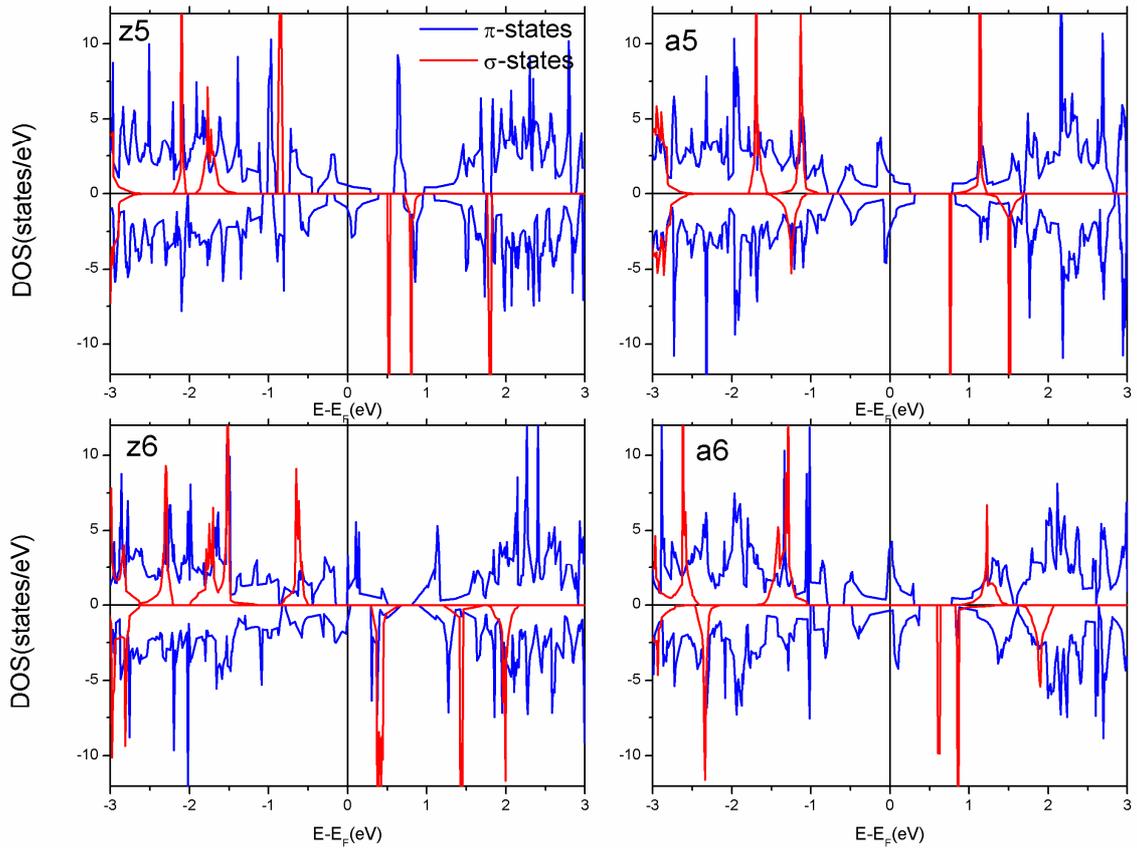

Figure 7. Partial density of states for: *z*5, *a*5, *z*6 and *a*6; indicating σ and π contributions. Blue and red lines represents π- and σ-states respectively.

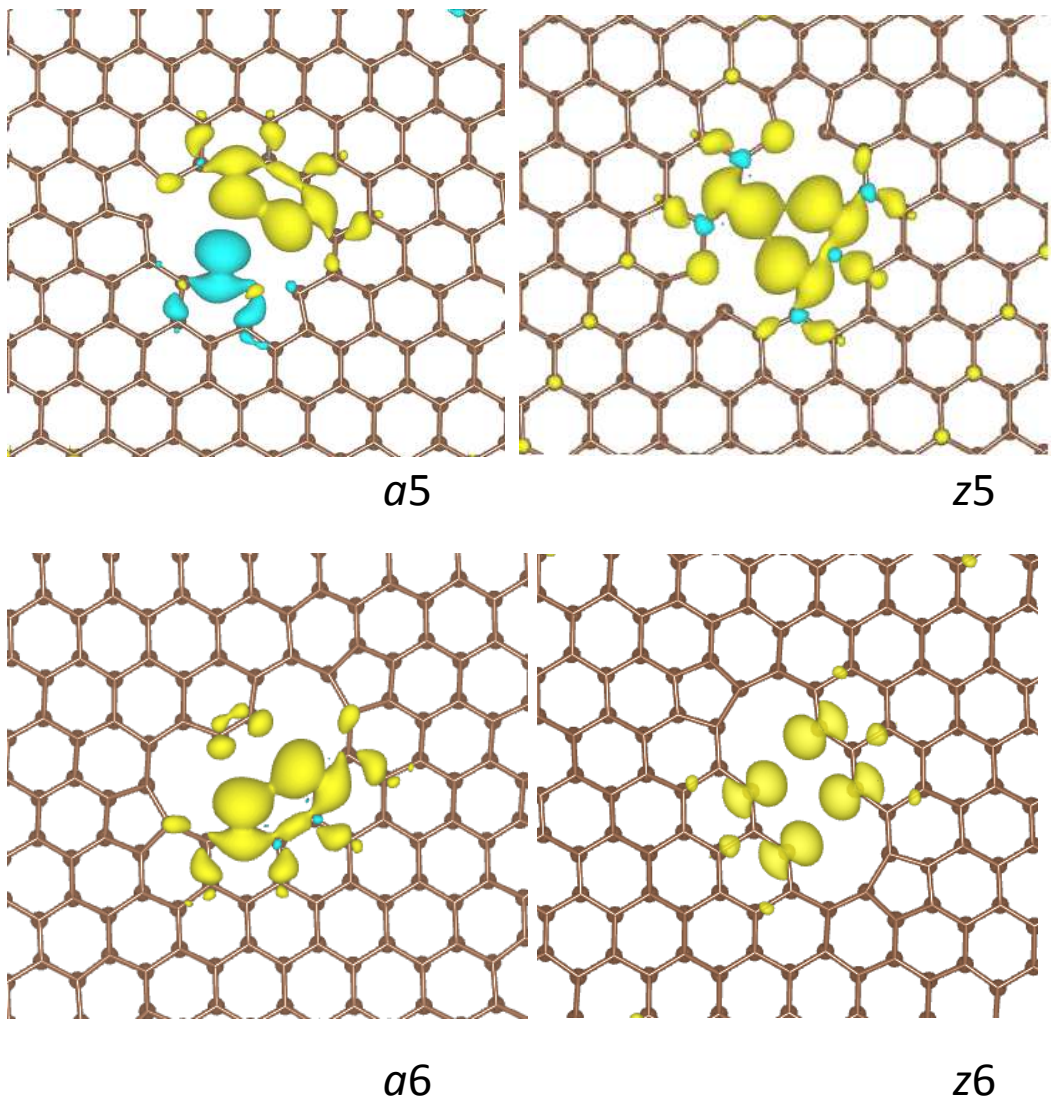

Figure 8. Spin-density maps for *a*5, *z*5, *a*6 and *z*6, indicating majority and minority contribution in yellow and cyan respectively. All the graphs were obtained for an iso-density of 0.01 e/Å$^3$.